\documentclass[12pt,a4paper,oneside]{article}
\usepackage{graphicx,sectsty,amssymb,amsmath,subfig,slashed}
\usepackage[linkcolor={blue},citecolor={red},colorlinks=true]{hyperref}
\usepackage[margin=2cm]{geometry}
\newcommand{\be}{\begin{equation}}
\newcommand{\ee}{\end{equation}}
\newcommand{\bea}{\begin{eqnarray}}
\newcommand{\eea}{\end{eqnarray}}

\newcommand{\cO}{\mathcal{O}}
\newcommand{\rg}{{r^g_{00}}}
\newcommand{\rbsp}{{\tilde r'_{bs}}}
\newcommand{\rsdp}{{\tilde r'_{sd}}}
\newcommand{\rsd}{{\tilde r_{sd}}}

\newcommand{\Reepsp}{{\rm Re}(\epsilon^\prime/\epsilon)}
\newcommand{\epsp}{{\epsilon^\prime/\epsilon_K}}
\newcommand{\epsK}{{\epsilon_K}}

\begin{document}

\title{\bf Combining Direct \& Indirect Kaon CP Violation to Constrain the Warped KK Scale}
\author{Oram Gedalia$^a$, Gino Isidori$^b$ and Gilad Perez$^a$ \vspace{6pt}\\
\fontsize{10}{16}\selectfont\textit{$^a$ Department of Particle Physics,
Weizmann Institute of Science, Rehovot 76100, Israel} \vspace{2pt}\\
\fontsize{10}{16}\selectfont\textit{$^b$ INFN, Laboratori Nazionali di Frascati,
Via E. Fermi 40 I-00044 Frascati, Italy}}
\date{}
\maketitle

\begin{abstract}
The Randall-Sundrum (RS) framework has a built in protection
against flavour violation, but still generically suffers from
little CP problems. The most stringent bound on flavour violation
is due to $\epsK$, which is inversely proportional to the
fundamental Yukawa scale. Hence the RS $\epsK$ problem can be
ameliorated by effectively increasing the Yukawa scale with a bulk
Higgs, as was recently observed in arXiv:0810.1016. We point out
that incorporating the constraint from $\epsp$, which is
proportional to the Yukawa scale, raises the lower bound on the KK
scale compared to previous analyses. The bound is conservatively
estimated to be 5.5~TeV, choosing the most favorable Higgs
profile, and 7.5~TeV in the two-site limit. Relaxing this bound
might require some form of RS flavour alignment. As a by-product
of our analysis, we also provide the leading order flavour
structure of the theory with a bulk Higgs.
\end{abstract}

\section{Introduction}

In generic RS models of a warped extra dimension with bulk fields,
the flavour puzzle is solved by the split fermion mechanism, where
the localization of fermions is determined based on their masses
and mixing angles~\cite{AS,RSoriginal}. Within the RS, this yields
extra protection against excess of flavour changing neutral
current (FCNC) processes in the form of RS-GIM~\cite{aps}. A
residual little CP problem is, however, still found in the form of
too large contributions to the neutron electric dipole
moment~\cite{aps} and sizable contributions to
$\epsK$~\cite{Bona:2007vi,Davidson:2007si,First,others} (see
also~\cite{RSflavor} for some related recent RS flavour studies).
Given an IR-localized Higgs field, a lower bound of ${\cal
O}(20)\,$TeV on the KK scale at leading order is
obtained~\cite{First,others}.

Recently in~\cite{Agashe:2008uz} it was pointed out, based on
matching the full RS set-up onto a two site model (originally
suggested in~\cite{Contino:2006nn}), that if the Higgs is in the
bulk and one-loop matching of the gauge coupling is included, the
KK scale can be lowered down to ${\cal O}(5)\,$TeV. An important
ingredient in that paper's analysis is the ability to raise the
overall size of the 5D down-type Yukawa coupling, $y^d$. The
resulting weaker bound is actually controlled by simultaneously
minimizing the contribution to $\epsK$, which effectively falls
like $1/(y^d)^2$, with the contribution to $b\to s\gamma$, which
grows like $(y^d)^2$. In this paper we point that a contribution
to $\epsp$, similar in structure to $b\to s\gamma$, actually
yields a much stronger constraint on the 5D Yukawa size, which
implies a strict conservative bound on the KK scale of 7.5~TeV in
the two site case. The bound is weakened to 5.5~TeV if one allows
the Higgs profile to saturate the AdS stability
bound~\cite{Breitenlohner:1982bm}. This is still beyond the LHC
reach~\cite{KKG}, and implies a rather severe little hierarchy
problem. We also show that UV-sensitive operators raise the bound
significantly, for instance in case the Higgs is localized on the
IR brane.

\section{Analysis}  \label{analysis}

\subsection{Flavour Structure with a Bulk Higgs}
In~\cite{Agashe:2008uz} it was pointed out that when the Higgs is
in the bulk, the light fermions can be made less composite while
still keeping their masses constant, and also the overall Yukawa
scale can be increased without violating the corresponding
perturbative bound. Both effects allow to ameliorate the RS
$\epsK$ problem. In our analysis below we carefully analyze the
flavour structure of the theory, allowing a rather general bulk
Higgs profile. In most of the past studies, the flavour structure of RS was
analyzed via the approximation that the Higgs and any relevant KK states are
localized on the IR brane, where a transparent spurion
structure can be formulated~\cite{aps}. Here we consider the
couplings by calculating full overlap integrals of wavefunctions,
and parametrize these corrections by appropriate functions of the
form:
\be
r=\frac{\textrm{wavefunctions overlap}}{\textrm{approximate
coupling on the IR}} \, .
\ee
This can be understood from some relevant sample terms in the 4D
effective Lagrangian~\cite{aps}:
\be \label{lagrangian}
\begin{split}
\mathcal{L}^{4D} \supset \sum_{i,j} & Y^d_{ij} {\cal H} \left[
\psi^{0 \dagger}_{Q_i} f_{Q_i} \psi^0_{d_j} f_{d_j} r^H_{00}
(\beta,c_{Q_i},c_{d_j})+ \sqrt{2} \sum_n \psi^{0 \dagger}_{Q_i}
f_{Q_i} \psi^n_{d_j} r^H_{0n} (\beta,c_{Q_i},c_{d_j}) \right. \\
& \left. + \sqrt{2} \sum_n \psi^{n \dagger}_{Q_i} \psi^0_{d_j}
f_{d_j} r^H_{n0} (\beta,c_{Q_i},c_{d_j}) + 2\sum_{n,m} \psi^{n
\dagger}_{Q_i} \psi^m_{d_j} r^H_{nm} (\beta,c_{Q_i},c_{d_j})
\right] \\ & +g_{s*} \sum_i G^1 \psi^{0 \dagger}_{i} \psi^{0}_{i}
\left( -\frac{1}{ k \pi R}+ f_i^2 \rg(c_i) \right) \, .
\end{split}
\ee
The term with square brackets is the coupling of the Higgs, ${\cal
H}$, to quarks of various zero/KK levels, $\psi^{0,n}$,
respectively. The other term is the coupling of zero-mode quarks
to the first KK gluon, $G^1$ and $i,j$ are flavour indices. For
simplicity we only present the down type quark couplings, where
$Q$ ($d$) stands for an SU(2) doublet (singlet) quark. The $f$'s
parametrize the values of SM quarks' profiles on the IR brane
(note that in the convention we follow, the value of KK fermions'
wavefunction on the IR brane is $\sqrt{2}$) and the $c$'s are
their bulk masses in units of $k$,
\begin{equation} \label{fc}
f(c) = \sqrt{\frac{1-2c}{1-(z_v/z_h)^{2c-1}}}\,.
\end{equation}
The coupling $Y^d_{ij}$ in Eq.~\eqref{lagrangian} is the 5D
anarchic down-type Yukawa matrix. We use $y^d$ to denote a generic
entry in $Y^d$ (in units of $\sqrt{k}$). Note that in comparison
to the notation of~\cite{Agashe:2008uz}, $Y_*^d=2y^d
r^H_{11}(\beta=1,c_1=0.55,c_2=0.55)$.

The KK decomposition for the Higgs is ${\cal H} (x,z)= \tilde{v}(
\beta, z) + \sum_n H^{(n)}(x) \phi_n(z)$~\cite{Agashe:2008uz},
where $\tilde{v}(\beta, z)$ is the Higgs VEV profile, which is
very close to the physical Higgs profile when $m_h \ll M_{KK}$
(here $\beta = \sqrt{4+\mu^2}$, with $\mu$ being the bulk Higgs
mass in units of $k$). This profile can be chosen to peak near the
IR brane:
\begin{equation}\label{bulkHiggs}
\tilde{v}( \beta, z) = v z_v
\sqrt{\frac{2(1+\beta)}{z_h^3(1-(z_h/z_v)^{2+2\beta})}}
\left(\frac{z}{z_v}\right)^{2+\beta}.
\end{equation}
For the purposes of the following discussion, the only important
parameter affecting the overlap corrections is $\beta$. The
$\beta=0$ case describes a Higgs maximally-spread into the bulk
(saturating the stability bound), while $\beta=1$ corresponds to
the two-site model considered in \cite{Agashe:2008uz}. For a
concrete comparison we take $\beta=1$, and add the case of the
weakest expected bound on the KK scale, which is obtained for
$\beta=0$.

Note that the case of an IR Higgs corresponds to setting the
$r^H$'s to unity. Full definitions and discussion of the
correction factors are presented in appendix~\ref{app:overlap}.

\subsection{RS Contributions to $\epsK$ and $b\to s\gamma$}

We start by considering the bound from $\epsK$. In this case
the largest contribution is generated by left-right effective
operators, and in particular by
\be
Q_4^K =\bar{d}_R^\alpha s_L^\alpha \bar{d}_L^\beta s_R^\beta \, .
\ee
In the RS framework the leading contribution to $C_4^K$ (the
effective coupling of $Q_4^K$) is generated by a tree-level
KK-gluon exchange. Up to $\cO(1)$ complex factors, this leads to
\begin{equation}
C_4^K ~\simeq~ \frac{g_{s*}^2}{M_{KK}^2} f_{Q_2} f_{Q_1} f_{d_2}
f_{d_1} \rg(c_{Q_2}) \rg(c_{d_2}) \approx
\frac{g_{s*}^2}{M_{KK}^2} \frac{ \lambda_d \lambda_s }{(y^d)^2}
\frac{\rg(c_{Q_2}) \rg(c_{d_2})}{r^H_{00}(\beta,c_{Q_1},c_{d_1})
r^H_{00}(\beta,c_{Q_2},c_{d_2})} \, . \label{eq:C4K}
\end{equation}
Here $M_{KK}$ is the scale of the first KK state, $g_{s*}$ is the
dimensionless 5D coupling of the gluon, and $\lambda_i$ is the SM
Yukawa coupling of the quark $i$ ($\lambda_i = m_{q_i}/v$,
$v\approx 174$~GeV). SM and 5D Yukawa couplings are connected by
the relation $\lambda_i \approx y^d f_{Q_i}
f_{d_i}r^H_{00}(\beta,c_{Q_i},c_{d_i})$. Eq.~(\ref{eq:C4K}) uses
the fact that the mixing angles of the rotation matrices from the
interaction basis to the mass basis are $f_{Q_i}/f_{Q_j}$ and
$f_{d_i}/f_{d_j}$ $(i\leq j)$ for the quark doublets and singlets,
respectively. We have verified numerically that the correction to
these relations due to the presence of a bulk Higgs in the
relevant range of parameters is a subleading effect. In principle,
terms proportional to $\rg(c_{Q_3,d_3})$ also contribute, with the
same $f_x's$ structure; however, since $\rg(c_{Q_3,d_3})<
\rg(c_{Q_2,d_2})$, the contribution shown in Eq.~(\ref{eq:C4K}) is
the dominant one.

The result in (\ref{eq:C4K}) is similar to the one given
in~\cite{Agashe:2008uz}. Taking into account the chirally-enhanced
$\langle K^0 | Q_4^K | \bar K^0 \rangle$ matrix element, assuming
an $\cO(1)$ CP violating phase\footnote{Note that here and below
we assume a single maximal CP violating phase. Given the fact that
each of the observables discussed by us actually receives
contributions from multiple independent terms, this is a
conservative assumption. A more reasonable approach might be to
estimate the sum of the different contributions via a ``random
walk'' approach, which will increase the amplitude by factor of
roughly $\sqrt{N/2}$, where $N$ is the number of independent
terms.} for $C_4^K$, requiring that the NP contribution to
$|\epsK|$ is 60\% of the experimental value~\cite{NMFV} and
evaluating the resulting suppression scale~\cite{Bona:2007vi} and
the quark masses~\cite{Xing:2007fb} at 5 TeV leads to
\begin{equation} \label{ek_bound1}
M_{KK} \gtrsim \frac{15 \, g_{s*}}{y^d} \, \mathrm{TeV} \, ,
\end{equation}
for $\beta=1$. The sources of difference from the result
of~\cite{Agashe:2008uz} are a correction to the overlap of the
quarks with the KK gluon and the 60\% saturation requirement. This
bound can be ameliorated by taking the Higgs to be maximally
spread into the bulk ($\beta=0$), which enhances its coupling to
the quarks (and raises the value of the mass corrections
$r^H_{00}$ included in the calculation above). The result in this
case is
\begin{equation} \label{ek_bound0}
M_{KK} \gtrsim \frac{8.5 \, g_{s*}}{y^d} \, \mathrm{TeV} \, .
\end{equation}

Next we consider $b \to s \gamma$. Here the largest contribution
is generated by the effective operator (we follow the conventions
of~\cite{Agashe:2008uz})
\be \label{bsg_operator}
Q_7^\prime = \frac{e m_b}{8\pi^2} \bar{b} \sigma^{\mu\nu}
F_{\mu\nu} (1+\gamma_5) s \, .
\ee
The effective coupling of $Q_7^\prime$ is generated in RS by a
loop diagram with a Higgs propagating in the loop~\cite{aps}, as
shown in Figure~\ref{fig:diagram}. The corresponding
Wilson coefficient, evaluated in
appendix~\ref{amp_calculation}, is
\begin{equation} \label{bsgamma}
C'_7 ~\approx~ \frac{1}{4 \lambda_b M_{KK}^2} f_{Q_3} (y^d)^3
f_{d_2} \rbsp\approx \frac{1}{4 M_{KK}^2} \frac{(y^d)^2
\lambda_s}{\lambda_b V_{ts}}
\frac{\rbsp}{r^H_{00}(\beta,c_{Q_2},c_{d_2})} \, ,
\end{equation}
where in the last equation we have used the relation
$f_{Q_2}/f_{Q_3} \approx V_{ts}$ between the left-handed profiles
($f_{Q_i}$) and the CKM matrix elements ($V_{ij}$)\footnote{We do
not distinguish here between gluon and quark KK masses, which only
slightly differ in the relevant range of parameters.}.
\begin{figure}[t]
\centering
\includegraphics[width=3.8In]{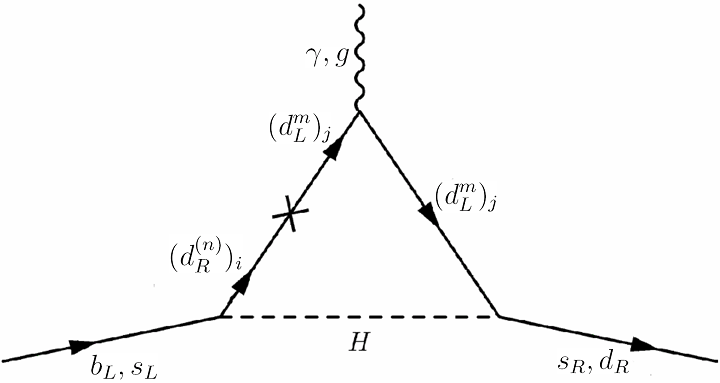}
\caption{Contribution to $b \rightarrow s\gamma$ and
$\epsp$ from Yukawa interactions.}
\label{fig:diagram}
\end{figure}
In Eq.~\eqref{bsgamma} we grouped together the overlap corrections to
the loop diagram under $\rbsp$, which contains contributions from
different quarks running in the loop. Under the assumption that
the Yukawa is anarchical, so that in the bulk interaction basis
the bulk masses are diagonal and the $c's$ are well
defined~\cite{aps}, $\rbsp$ is:
\begin{equation}
\rbsp \approx \sum_{i,j} r^H_{0n}(\beta,c_{Q_3},c_{d_i}) \left[
{2r^H_{n^-m^-}(\beta,c_{Q_j},c_{d_i})-\frac{1}{3}r^H_{nm}(\beta,c_{Q_j},c_{d_i})}
\right]^\dagger r^H_{m0}(\beta,c_{Q_j},c_{d_2}) \,,\label{rbsp}
\end{equation}
where $i,j$ are flavour indices and $m,n$ are the KK levels of the
fermions in the loop, and we only consider the first KK state,
since at one-loop the above contribution is finite (or at most
log-divergent)~\cite{aps,Agashe:2006iy,PR-neutrino}. Following the
analysis in~\cite{Agashe:2008uz} (allowing 20\% departure from the
SM value of $\mathcal{B}(B \to X_s \gamma)$) and using the values
of the quark Yukawa couplings at 5 TeV~\cite{Xing:2007fb}, we
obtain for $\beta=1$ by requiring $C'_7(5\textrm{
TeV})/C^{\mathrm{SM}}_7(M_W)<1.4$:
\begin{equation}
\label{bsg_bound1}
M_{KK} \gtrsim 0.4 \, y^d \, \mathrm{TeV} \, .
\end{equation}
This result is \emph{lower} than the corresponding bound reported
in~\cite{Agashe:2008uz}, because of the overlap corrections
considered here. Contrary to $\epsK$, the case of $\beta=0$
yields the same bound as for $\beta=1$. As discussed
in~\cite{aps,Davidson:2007si,Agashe:2008uz}, the $b \to s \gamma$
operator with opposite chirality leads to a weaker condition.

In~\cite{Agashe:2008uz} the constraints from $\epsK$ and $b \to s
\gamma$ in (\ref{ek_bound1}) and (\ref{bsg_bound1}) are combined
to evaluate the value of $Y_*^d$ that minimizes the lower bound on
$M_{KK}$, and the coupling $g_{s*}$ is matched to the measured 4D
coupling at one-loop, resulting in $g_{s*}\approx 3$
($g_{s*}\approx 6$ at the tree-level). Using this analysis for our
results naively gives a bound of about 4 TeV, but the value of
$y^d$ is above the perturbativity bound
($4\pi/\sqrt{N_{KK}}$)~\cite{Agashe:2008uz}. Taking $y^d$ equal to
this bound with the minimal conceivable value $N_{KK}=2$ gives for
$\beta=1$:
\begin{equation}
M_{KK} \gtrsim 5 \, \mathrm{TeV} \, .
\end{equation}

\subsection{The Constraint of $\epsp$}

As anticipated in the introduction, here we show that
the bound in (\ref{bsg_bound1}) becomes substantially
stronger after we include an additional constraint from $\Reepsp$,
the direct CP violating observable of the $K^0 \to 2\pi$ system.
As pointed out in~\cite{Davidson:2007si},
the constraint following from the contribution of
the chromomagnetic operator to $\Reepsp$  is similar in structure
to the $b \to s \gamma$ one, but is numerically more stringent.

Before analysing the extra contribution to $\Reepsp$
generated in the RS framework, it is worth to briefly
recall the experimental status of this observable
and its prediction within the SM:
\begin{itemize}
\item After a series of measurements by the KTeV and the NA48
collaborations, the present experimental world average is
$\Reepsp_{\rm exp} =(1.65 \pm 0.26) \times 10^{-3}$~\cite{PDG}.
\item $\Reepsp_{\rm SM}$ is dominated by the contributions of two
operators: the electroweak penguin, contributing to ${\rm
Im}(A_{2})= {\rm Im}[{\mathcal A}(K\to (2 \pi)_{I=2})]$, and the
QCD penguin ($Q_6$ in the standard notations), contributing to
${\rm Im}(A_{0})$. The destructive interference of these two
contributions is one of the reasons why it is difficult to obtain
a precise estimate of $\Reepsp_{\rm SM}$. The negative
contribution to $\Reepsp_{\rm SM}$ generated by the electroweak
penguins is estimated with $20\%$--$30\%$ errors, both on the
lattice and using analytic methods~\cite{Cirigliano:2004xp}. On
the other hand, the chiral structure of $Q_6$ and the sizable
final-state interactions in the $(2\pi)_{I=0}$ channel prevent, at
present, a reliable estimate of this matrix element on the
lattice~\cite{Lee:2006cm}.\footnote{~The difficulty in estimating
$\langle 2\pi | Q_6 |K^0 \rangle$ on the lattice is confirmed by
the difficulty of reproducing the experimental value of Re($A_0$)
on the lattice. The latter is affected by similar problems (in
particular the large final-state interactions), but it is free
from new-physics contaminations. In particular, lattice estimates
tend to underestimate Re($A_0$): this provide a qualitative
understanding of why lattice estimates of $\Reepsp_{\rm SM}$ are
typically smaller (or even negative) compared to the analytic
ones. } Recent estimates based on analytic
methods~\cite{Bertolini:2000dy,Pich:2004ee} lead to values of
$\Reepsp_{\rm SM}$ in good agreement with $\Reepsp_{\rm exp}$,
with errors ranging from $30\%$ to $50\%$. As a conservative
approach, in the following we assume a $100\%$ error, or $0 <
\Reepsp_{\rm SM} < 3.3 \times 10^{-3}$, consistently with the
conservative range suggested in~\cite{Buras:2001au}.
\end{itemize}

The potentially large new contribution to $\Reepsp$ in the RS
framework is induced by the two effective chromomagnetic operators
\be
\label{chromo_operator}
Q_G = g_s H^\dagger \bar{s}_R \sigma^{\mu\nu} T^a G^a_{\mu\nu}
d_L~, \qquad Q_G^\prime = g_s H \bar{s}_L \sigma^{\mu\nu} T^a
G^a_{\mu\nu} d_R~.
\ee
Similar to the $b \to s \gamma$ case, these are generated by the
Higgs-mediated one-loop amplitude in Fig.~\ref{fig:diagram}. The
coefficients, evaluated in appendix~\ref{amp_calculation}, are
\be
\begin{split}
C_G  &\approx \frac{3}{16\pi^2 M_{KK}^2} f_{Q_1} (y^d)^3 f_{d_2}
\rsd \approx \frac{3}{16\pi^2 M_{KK}^2} (y^d)^2 \lambda_s V_{us}
\frac{\rsd}{r^H_{00}(\beta,c_{Q_2},c_{d_2})} ~, \\
C^\prime_G &\approx \frac{3}{16\pi^2 M_{KK}^2} f_{Q_2} (y^d)^3
f_{d_1} \rsdp \approx \frac{3}{16\pi^2 M_{KK}^2} \frac{(y^d)^2
\lambda_d}{ V_{us} } \frac{\rsdp}{r^H_{00}(\beta,c_{Q_1},c_{d_1})}
~, \label{chromomagnetic}
\end{split}
\ee
where $\rsd$ and $\rsdp$ group the overlap corrections (see
Eq.~\eqref{chromo_coefficient}). Defining
\begin{equation}
\delta_{\epsilon^\prime} = \frac{ \Reepsp_{\rm RS} - \Reepsp_{\rm SM} }{ \Reepsp_{\rm exp} }~,
\end{equation}
we obtain
\be
\delta_{\epsilon^\prime} = \frac{\omega \langle (2\pi)_{I=0} | \lambda_s Q_G | K^0 \rangle  }{
\sqrt{2}  {\rm Re} A_0 \Reepsp_{\rm exp} |\epsilon|_{\rm exp} }
\left[ \frac{{\rm Im}(C_G -C^\prime_G)}{\lambda_s} \right]
\approx (58~{\rm TeV})^2 B_G \left[ \frac{{\rm Im}(C_G -C^\prime_G)}{\lambda_s} \right]~,
\ee
where $B_G$ is the hadronic bag-parameter defined
by~\cite{Buras:1999da}\footnote{~Here we adopt the notation
of~\cite{Buras:1999da}, where $F_\pi =131$~MeV and the
$K^0 \to (2\pi)_{I}$ amplitudes are normalised such that ${\rm
Re}(A_0) =  3.3 \times 10^{-4}$~MeV (note that the analog of
Eq.~(\ref{eq:BG}) reported in~\cite{Davidson:2007si} has a missing
factor 1/2). Additional numerical inputs are $\omega = |A_2/A_0|= 0.045$,
and $|\epsilon|_{\rm exp} = 2.23\times 10^{-3}$.}
\be
\langle 2\pi_{I=0} | \lambda_s Q_G | K^0 \rangle =
\sqrt{\frac{3}{2}} \frac{11}{4} \frac{ m_\pi^2 m_K^2 }{ F_\pi} B_G~.
\label{eq:BG}
\ee
The value $B_G=1$ corresponds to the estimate of this hadronic
matrix element in the chiral quark model and to the first order in
the chiral expansion~\cite{Bertolini:1994qk}. A similar numerical
value is also obtained using different hadronization
techniques~\cite{He:2000xa}. The hadronic matrix element of the
chromomagnetic operator is affected by the same difficulties
appearing in  $\langle 2\pi | Q_6 |K^0 \rangle$: beyond the
lowest-order in the chiral expansion we expect large positive
corrections from final-state interactions. Moreover, as pointed
out in~\cite{Buras:1999da}, higher order chiral corrections should
remove the accidental  $m_\pi^2$ suppression in Eq.~(\ref{eq:BG}).
Therefore the estimate of $\delta_{\epsilon^\prime}$ obtained
with $B_G=1$ can be considered as a conservative lower bound. The
leading QCD corrections in running down the Wilson coefficients
from the high scale ($\sim 5$ TeV) down to $\sim 1$ GeV are taken
into account by the running of the (4D) Yukawa couplings (the
residual effect is smaller than 15\%~\cite{Buras:1999da}). As a
result, the ratios $C^{(\prime)}_G/\lambda_s$ and the matrix
element in (\ref{eq:BG}) are, to a good approximation,
scale-independent quantities.

Assuming $\cO(1)$ CP violating phases for $C_G$ and $C^\prime_G$,
barring accidental cancellations among these two terms and
imposing $|\delta_{\epsilon^\prime}| < 1$, leads to
\begin{equation}
M_{KK} \gtrsim 1.3 \, y^d \, \mathrm{TeV} \label{eq:epsp_bound1}
\end{equation}
for $\beta=1$, and
\begin{equation}
M_{KK} \gtrsim 1.2 \, y^d \, \mathrm{TeV} \label{eq:epsp_bound0}
\end{equation}
for $\beta=0$.

\begin{figure}[t]
\centering
\subfloat[]{
\includegraphics[width=3.2In]{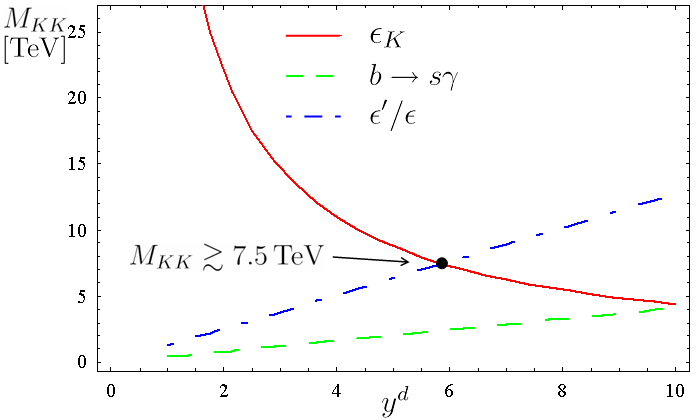}} \hspace{0.1In}
\subfloat[]{\includegraphics[width=3.2In]{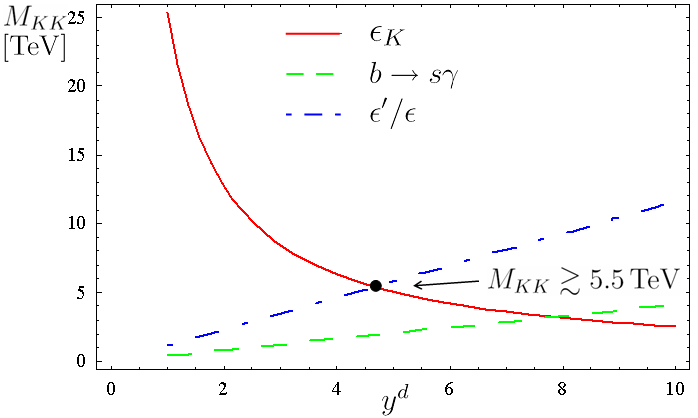}}
\caption{The combination of the bounds on $M_{KK}$, as a function
of $y^d$, for (a)~$\beta=1$, which corresponds to the two site
model; (b)~$\beta=0$, which corresponds to the weakest bound.}
\label{fig:bound}
\end{figure}

The constraint in Eq.~\eqref{eq:epsp_bound1} is substantially
stronger with respect to the one in Eq.~\eqref{bsg_bound1} (note
also that the former only depends on the down-type Yukawa, while
the $b \rightarrow s\gamma$ amplitude, which is dominated by a
charged Higgs contribution, implicitly depends on the up-type
Yukawa, too). When combined with Eq.~\eqref{ek_bound1}, the
overall bound is obtained for $y^d \approx 5.9$ (assuming $g_{s*}
\approx 3$) and is
\begin{equation}
M_{KK} \gtrsim 7.5 \, \mathrm{TeV} \, .
\end{equation}
The lowest possible bound comes from combining
Eq.~\eqref{ek_bound0} and Eq.~\eqref{eq:epsp_bound0}:
\be \label{final_ideal}
M_{KK} \gtrsim 5.5 \, \mathrm{TeV} \, .
\ee
The combination of the bounds is shown in Fig.~\ref{fig:bound}.

The reason why the $\epsp$ constraint is substantially stronger
than $b\to s\gamma$ one (more than a factor of 10 at the amplitude
level) can be understood by comparing the parametric dependence
from quark masses and CKM factors of the $LR$ (chromo-)magnetic
operators in the RS framework~\cite{Davidson:2007si}:
\begin{equation}
\frac{ A( b_{L} \to s_{R} \gamma(g))_{KK} }{ A( b_R \to s_L \gamma(g) )_{\rm SM} }
\propto \frac{m_s}{m_b |V_{ts}|^2} \sim \frac{1}{|V_{us}|^2}  \qquad {\rm vs. } \qquad
\frac{ A( s_R \to d_L  \gamma(g))_{KK} }{ A( s_R \to d_L  \gamma(g))_{\rm SM} }
\propto \frac{ |V_{us}| }{ |V_{ts}^* V_{td}| } \sim \frac{1}{|V_{us}|^{4}}~. \qquad
\end{equation}
In principle, the large enhancement of the $s_R \to d_L$ magnetic
transitions in the RS framework occurs also in the short-distance
component of the $s\to d\gamma$ amplitude. In most $K$ decays this
amplitude is unmeasurable, being obscured by long-distance
contributions. The only case were it could be detected is the rare
decay $K_L \to \pi^0 e^+e^-$~\cite{Buras:1999da,Buchalla:2003sj}.
However, present experimental data on this decay mode are still
far from the SM level~\cite{PDG}, and the corresponding bound on
$M_{KK}$ is not competitive even with Eq.~\eqref{bsg_bound1}
(see~\cite{Blanke:2008yr} for a study of rare decys in the context
of RS).

\subsection{The Case of an IR-Brane Higgs with UV Sensitive Operators}
As already mentioned above, there are cases where the leading
contributions are from UV sensitive operators. This is expected
since the 5D theory is non-renormalizable with negative mass
dimension operators. For instance, when the Higgs is localized on
the IR brane, the above one-loop is divergent, and a counter-term
in the form of a higher dimensional IR operator is
included~\cite{aps,Agashe:2006iy,PR-neutrino}. We can derive a
bound on the corresponding cutoff of the theory required to
satisfy the constraints from $b\to s\gamma$ and $\epsp$ (similarly
to the $\mu\to e\gamma$ case~\cite{PR-neutrino}). The effective
operator on the IR can be written as
\begin{equation}
(O^{IR}_{x})_{bs, sd}= \frac{g^x_5
Q_d}{\left(\Lambda^{bs,sd}_{IR}\right)^2}   \,Y_d\, H
\bar Q^i \sigma_{\mu \nu} d^j K^{\mu \nu}\,,
\end{equation}
where for the case of $b\to s \gamma$ ($\epsp$), $x=1$ ($x=3$),
$i,j=3,2$ ($i,j=2,1$ or $i,j=1,2$), $Q_d=1/3$ ($Q_d=1$) and
$K^{\mu\nu}=F^{\mu \nu}$ ($K^{\mu\nu}=G^{\mu \nu}$), the
electromagnetic (gluon) field strength. The above expression is
simplified when we switch from 5D fields to canonically normalized
4D ones. This cutoff operator can be rewritten (in terms of the
zero-modes and 4D couplings) on the IR brane as~\cite{PR-neutrino}
\be
O^{IR}_{bs}\sim   \frac{e}{6\left(\Lambda^{bs}_{IR}\right)^2}\,
\frac{m_s}{V_{ts}} \, F^{\mu\nu} \bar b \,\sigma_{\mu \nu}
(1+\gamma_5)s\,,\,\, \ \ \ \ O^{IR}_{sd}\sim
\frac{g}{2\left(\Lambda^{sd}_{IR}\right)^2}\, \frac{m_d}{V_{us}}
\, G^{\mu\nu} \bar s \,\sigma_{\mu \nu} d_R\,,
\ee
where we have replaced the Higgs by its vev and, for simplicity,
we only consider the case where $i,j=2,1$ for $\epsp$.
Repeating the above analysis, we find the following bounds for
$\Lambda^{bs,sd}_{IR}$
\be
\Lambda^{bs}_{IR}\gtrsim 8  {\rm\, TeV}~, \qquad
\Lambda^{sd}_{IR}\gtrsim 20 {\rm\, TeV}\, .
\ee

\section{Conclusion}

Generic flavour models within the RS framework provide an elegant
explanation of the fermion mass hierarchy; however, the resulting
suppression of FCNC processes might not be enough. In this paper
we have shown that the constraints stemming from $\epsp$ yield a
lower bound of at least $\sim 5.5\,$TeV on the KK scale.

The numerical value of this bound is highly insensitive to the
precise value of the quark bulk mass parameters. Moreover, there
are various reasons to consider this result as very conservative:
\begin{itemize}
\item the bag parameter of the chromo-magnetic operator
      is likely to exceed the reference
      value we have adopted ($B_G=1$), and we have allowed
      the RS contribution to saturate the experimental
      measurement of $\epsp$;
\item we have only considered the first KK level of the quarks and the
      zero-mode of the Higgs;
\item since the value of $y^d$ is close to
      the perturbativity bound, the contribution from higher loops is
      probably not negligible and, a piori, does not need not to be
      suppressed by $r^H_{1^- 1^-}$;
\item the final bound is obtained with an ``ideal'' Higgs profile.
\end{itemize}
A more realistic evaluation should result in a stronger bound.

The bound thus obtained is stronger than the one derived from
electroweak precision tests, and induces a rather severe little
hierarchy problem. If taken a face value, it also implies that a
direct LHC discovery of the relevant degrees of freedom is
unlikely. This motivates the search for alternative solutions of
the residual RS flavour problem, such as the inclusion of some
form of flavour alignment of the fundamental down-type
sector~\cite{PR-neutrino,Align}.

\section*{Acknowledgements}

We are grateful to Kaustubh Agashe for very valuable
discussions and comments on the manuscript. We also
thank Uli Haisch and Marek Karliner for useful discussions,
and Yossi Nir for comments on the manuscript.
The work of GP is supported by the Peter and Patricia
Gruber Award. The work of GI is supported by the EU under contract
MTRN-CT-2006-035482 ({\em Flavianet}).

\appendix
\section{Overlap Corrections} \label{app:overlap}
A common approach is to use the values of fermion zero-modes on
the IR brane, parameterized by $f$'s, to evaluate their coupling
to the Higgs. This is exact in case the Higgs is localized on the
IR brane, but it is only an approximation for a bulk Higgs. A
similar approximation is used in the coupling of fermions to the
KK gluon, which is concentrated near the IR brane. Since we are
trying to constrain the RS flavour model, a more careful treatment
is required. In this appendix we estimate the corrections to these
approximations by calculating full overlap integrals of the
wavefunctions. For consistency, we follow the conventions
of~\cite{Agashe:2008uz}, and use the definitions given in
appendix~E of that paper.

The correction factor for the overlap of the Higgs zero-mode with
two zero-mode fermions is given by
\be \label{rh00}
\begin{split}
r^H_{00}(\beta,c_L,c_R) & \equiv \frac{\int_{z_h}^{z_v} dz
(z_h/z)^5 \tilde{v}(\beta,z) \chi_0(c_L,z) \chi_0(c_R,z)
/(v\sqrt{k})}{\chi_0(c_L,z_v) \chi_0(c_R,z_v) z_h^4/z_v^3}=
\\ &=\frac{(1-e^{-(2+\beta-c_L-c_R)k \pi
R})\sqrt{2(1+\beta)}}{2+\beta-c_L-c_R} \simeq
\frac{\sqrt{2(1+\beta)}}{2+\beta-c_L-c_R} \, ,
\end{split}
\ee
where $\tilde{v}(\beta,z)$ is the Higgs zero-mode wavefunction
defined in Eq.~\eqref{bulkHiggs} and $\chi_0(c,z)$ is the fermion
zero-mode wavefunction (see also Eq.~\eqref{fc}):
\be
\chi_0(c,z)=\frac{f(c)}{\sqrt{z_h}} \left(\frac{z_h}{z_v}
\right)^{1/2-c} \left( \frac{z}{z_h} \right)^{2-c} \, .
\ee
The last approximate equality in Eq.~\eqref{rh00} is valid to a
very good accuracy for $2+\beta>c_L+c_R$ (which is related to the
``switching'' behavior discussed in \cite{Agashe:2008fe}). This
result can be conveniently factorized to some approximation by
\be
r^H_{00} \approx \frac{4 \sqrt{2+2\beta}}{2+\beta} \frac{1}{2-c_L}
\frac{1}{2-c_R} \, ,
\ee
which is valid for $\beta \lesssim 5$ and when at least one of the
$c$'s is close to 0.

Similarly, we define the correction factors for the overlap of the
Higgs with a zero-mode fermion and a KK fermion, and with two KK
fermions, respectively
\be
\begin{split}
r^H_{01}(\beta,c_0,c_1) & \equiv \frac{\int_{z_h}^{z_v} dz
(z_h/z)^5 \tilde{v}(\beta,z) \chi_0(c_0,z) \chi_1(c_1,z)
/(v\sqrt{k})}{\chi_0(c_0,z_v) \chi_1(c_1,z_v) z_h^4/z_v^3} \, , \\
r^H_{11}(\beta,c_1,c_2) & \equiv \frac{\int_{z_h}^{z_v} dz
(z_h/z)^5 \tilde{v}(\beta,z) \chi_1(c_1,z) \chi_1(c_2,z)
/(v\sqrt{k})}{\chi_1(c_1,z_v) \chi_1(c_2,z_v) z_h^4/z_v^3} \, .
\end{split}
\ee
For simplicity, we focus only on the first KK state of the
fermions $\chi_1(c,z)$, defined by
\be \label{kkfermion1}
\chi_1(c,z)=\frac{1}{N_1 \sqrt{\pi R}} \left( \frac{z}{z_h}
\right)^{5/2} \left[ J_\alpha (m_1 z)+b_\alpha (m_1) Y_\alpha (m_1
z) \right] \, ,
\ee
with
\be \label{kkfermion2}
\begin{split}
-&b_\alpha (m_1) =\frac{J_{\alpha-1}(m_1 z_h)}{Y_{\alpha-1}(m_1
z_h)} =\frac{J_{\alpha-1}(m_1 z_v)}{Y_{\alpha-1}(m_1 z_v)} \, , \\
N_1 &=\frac{1}{2 \pi R} \left\{ z_v^2 \left[ J_\alpha (m_1
z_v)+b_\alpha (m_1) Y_\alpha (m_1 z_v) \right]^2- z_h^2 \left[
J_\alpha (m_1 z_h)+b_\alpha (m_1) Y_\alpha (m_1 z_h) \right]^2
\right\} \, ,
\end{split}
\ee
and $\alpha \equiv |c+1/2|$. $r^H_{01}(\beta,c_0,c_1)$ can be
computed analytically, but the result is very complicated, while
for $r^H_{11}(\beta,c_1,c_2)$ we could not find an analytic
solution. Reasonable polynomial fits are given by
\be
\begin{split}
r^H_{01}(\beta,c_0,c_1) & \approx 0.41+0.39 c_1-0.04 \beta c_1
+0.15 c_0 c_1 +0.10 c_1^2 -0.24 c_1^3 \, , \\
r^H_{11}(\beta,c_1,c_2) & \approx 0.32+0.06 \beta -0.01 \beta^2
+0.08 (c_1+c_2) +0.20 c_1 c_2 -0.08 c_1 c_2 (c_1+c_2) \, ,
\end{split}
\ee
The first one is valid to an accuracy of about 10\% for
$\beta\lesssim 10$ (and breaks down in the region where $\beta
\sim 0$ and $|c_1-c_0|\gtrsim 1$) and the second for
$\beta\lesssim 6$.

The wavefunction defined in Eqs.~\eqref{kkfermion1}
and~\eqref{kkfermion2} describes a KK fermion with the same
chirality as the zero-mode fermion, that is, with $\{++\}$
boundary conditions. There is also a KK fermion with opposite
chirality ($\{--\}$ boundary conditions) $\tilde{\chi}_1(c,z)$,
defined in the same way as in Eq.~\eqref{kkfermion1}, but with
\be
\begin{split}
-&\tilde{b}_\alpha (m_1) =\frac{J_{\alpha}(m_1
z_h)}{Y_{\alpha}(m_1
z_h)} =\frac{J_{\alpha}(m_1 z_v)}{Y_{\alpha}(m_1 z_v)} \, , \\
\tilde{N}_1 &=\frac{1}{2 \pi R} \left\{ z_v^2 \left[ J_{\alpha-1}
(m_1 z_v)+\tilde{b}_\alpha (m_1) Y_{\alpha-1} (m_1 z_v) \right]^2-
z_h^2 \left[ J_{\alpha-1} (m_1 z_h)+\tilde{b}_\alpha (m_1)
Y_{\alpha-1} (m_1 z_h) \right]^2 \right\} \, ,
\end{split}
\ee
and the replacement $c \rightarrow -c$. Regarding the overlap of
two KK fermions with the Higgs, we actually mostly use the
opposite chirality states
\be
r^H_{1^- 1^-}(\beta,c_1,c_2) \equiv \frac{\int_{z_h}^{z_v} dz
(z_h/z)^5 \tilde{v}(\beta,z) \tilde{\chi}_1(c_1,z)
\tilde{\chi}_1(c_2,z) /(v\sqrt{k})}{\tilde{\chi}_1(c_1,z_v)
\tilde{\chi}_1(c_2,z_v) z_h^4/z_v^3} \, ,
\ee
rather than $r^H_{11}(\beta,c_1,c_2)$. A polynomial fit to
$r^H_{1^- 1^-}$ is given by
\be
r^H_{1^- 1^-}(\beta,c_1,c_2) \approx 0.34-0.06 \beta +0.01\beta^2
+0.05 (c_1+c_2) -0.06 (c_1^3+c_2^3) \, ,
\ee
valid to $\beta \lesssim 5$.

The last correction factor we use is for the coupling of a KK
gluon with two zero-mode fermions
\be
\rg(c) \equiv \frac{\int_{z_h}^{z_v} dz (z_h/z)^4
(f_1(z)-f_1(z_h)) \chi_0^2(c,z)/\sqrt{k}}{\chi_0^2(c,z_v)
z_h^4/z_v^3} \, ,
\ee
where $f_1(z)$ is the wavefunction of the first KK gluon, and we
subtract its value on the UV brane because it represents the
flavour-universal part. This formula has a useful approximation,
obtained by neglecting the Y-type Bessel function in the KK gluon
wavefunction \cite{First,kkgluon}:
\be
\rg(c)=\frac{\sqrt{2}}{J_1(x_1)} \int_0^1 x^{1-2c} J_1(x_1 x) dx
\approx \frac{\sqrt{2}}{J_1(x_1)} \frac{0.7}{6-4c} \left(
1+e^{c/2} \right) \, ,
\ee
with $x_1 \approx 2.4$ being the first root of the Bessel function
$J_0(x_1)=0$.

In order to calculate realistic correction factors for the
operators considered above, we employ the following procedure.
First we choose values for three basic parameters: $\beta$, $y^d$
and $c_{Q_3}$ (the bulk mass parameter of the third generation
left-handed quark doublet). The masses of the other left-handed
doublets are obtained by the relation $f_{Q_i}/f_{Q_j} \sim
V_{ij}$, and the masses of the right-handed quarks are extracted
from $\lambda_i \simeq y^d f_{Q_i}
f_{d_i}r^H_{00}(\beta,c_{Q_i},c_{d_i})$. Finally, the relevant
correction factors for each operator are computed together (note
that $r^H_{00}$ appears only when the mass relation is used).

Using this procedure, it was found that the corrections are
actually quite insensitive to the value of $c_{Q_3}$ in the range
of $0-0.6$ and to $y^d$ around the value that minimizes the
overall bound, which makes this analysis robust (although for a
third generation quark running in the loop the correction is a bit
more sensitive to $c_{Q_3}$). The only important parameter is
$\beta$, as can be expected (e.g.~the overlap of two light
fermions with a bulk Higgs is quite different than with an IR
brane-localized Higgs).

The main result is that for the operators responsible for $b
\rightarrow s \gamma$ and $\epsp$, the coefficients are reduced by
about an order of magnitude (for $\beta=0,1$), lowering the bound
on the KK scale. This primarily stems from the correction
$r^H_{1^- 1^-}$ and the mass correction $r^H_{00}$ ($r^H_{1^-
1^-}$ is always smaller than $r^H_{11}$, so the inclusion of the
former in the dipole operators reduces their contribution relative
to what might be naively expected). The $\epsK$ contribution is
actually raised for $\beta=1$, as a result of the KK gluon overlap
correction $\rg$.

An important comment is in order. The bulk Higgs zero-mode
wavefunction is usually obtained by adding a bulk mass for the
Higgs and kinetic terms on \emph{both} branes (otherwise the
zero-mode vanishes by boundary conditions). Hence, There is no
smooth limit (e.g.~$\beta \rightarrow \infty$) in which the bulk
Higgs zero-mode wavefunction corresponds to an IR-localized Higgs.

\section{One-Loop Coefficients of the Dipole Operators}
\label{amp_calculation}

\subsection{The One-Loop Integral}

Up to the overall coupling dictated by the flavour structure and
the wave-function overlaps, the amplitude for the diagram in
Fig.~\ref{fig:diagram} with an external gluon line (including only
the contributions from the down-type flavour sector) is
\be
\begin{split}
iA(s\rightarrow d \, g)&=\int
\frac{d^4k}{(2\pi)^4}\overline{u}(p')
\frac{(\hat{\slashed{p}}'+M_{KK}^{(i)})}{\hat{p}^{'2}-M_{KK}^{(i),2}}
(g_s\gamma^\mu t^a G^a_\mu)\frac{i(\hat{\slashed{p}}
+M_{KK}^{(i)})}{\hat{p}^2-M_{KK}^{(i),2}} (1\pm\gamma_5) u(p)
\cdot\frac{1}{k^2-m_H^2} \\
&= \int \frac{d^4k}{(2\pi)^4} \overline{u}(p')\left[g_s t^a
G^a_\mu M_{KK}^{(i)} \frac{\hat{\slashed{p}}'\gamma^\mu
+\gamma^\mu \hat{\slashed{p}}}
{(\hat{p}^{'2}-M_{KK}^{(i),2})(\hat{p}^2-M_{KK}^{(i),2})(k^2-m_H^2)}\right]
(1\pm\gamma_5) u(p) \, ,
\end{split}
\ee
where $\hat{p}^{(\prime)}=p^{(\prime)}+k$. Neglecting the Higgs
mass, this leads to
\be \label{amp_no_flavor}
\begin{split}
A&(s\rightarrow d\, g)= \\ &=g_s t^a G^a_\mu M_{KK} \int \frac{d^4
l}{(2\pi)^4}\int_0^1dx \int_0^{1-x} dy \overline{u}(p')
\frac{-x(p^\mu +i\sigma^{\mu \nu} p_\nu)-y(p'^\mu-i\sigma^{\mu
\nu}p'_\nu)} {\left[l^2-  M_{KK}^{(i),2}(x+y) \right]^3} (1\pm\gamma_5) u(p) \\
& =\frac{g_s t^a G^a_\mu}{4(4 \pi)^2 M_{KK}} \overline{u}(p')
\sigma^{\mu\nu}q_\nu (1\pm\gamma_5) u(p) \, ,
\end{split}
\ee
where $q \equiv p'-p$, we have used the equations of motion on the
external spinors, and we have set the KK mass equal to the value
of the first KK state.

\subsection{Diagonalization of the Quark Mass Matrix}
\label{mass_diag}

In order to compute the overall coupling of the loop amplitude, we
need to address the diagonalization of the quark mass matrix. In
general, a zero-mode quark in the interaction basis mixes with the
KK states. Restricting the discussion to the first KK level, there
are two different states: one with the ``right'' chirality
($\{++\}$ boundary conditions, similar to the zero-mode), the
other with ``wrong'' chirality ($\{--\}$ boundary conditions,
projecting out the zero-mode). The actual contribution to a
measurable quantity is then calculated after the mixing matrix is
diagonalized to the mass basis (see appendix B
in~\cite{Agashe:2008uz} and~\cite{Buras:2009ka} for similar
analyses).

For simplicity, we consider a one generation case,
so the mass matrix $M_q$ is given by
\bea
\begin{pmatrix} \bar Q^{(0)}& \bar d^{(1)}_L &\bar Q^{(1)}_L
\end{pmatrix}
M_q
\begin{pmatrix}  d^{(0)}\\ Q^{(1)}_R\\   d^{(1)}_R
\end{pmatrix} \, , \qquad M_q= M_{KK}
\begin{pmatrix} x f_Q f_d r^H_{00} &0 & \sqrt{2} x f_Q r^H_{01} \\
0 & 2x r^H_{1^- 1^-} & 1  \\ \sqrt{2} x f_d r^H_{10} &1 & 2x
r^H_{11}
\end{pmatrix} \, ,
\eea
with $x \equiv v y^d/M_{KK}$. $M_q$ is diagonalized to first order
in $x$ by a bi-unitary transformation:
\be
M_q^{\mathrm{mass}}=O_L^{\dagger} M_q O_R=M_{KK} \times
\mathrm{diag}(x f_Q f_d r^H_{00},1+x(r^H_{11}+r^H_{1^-
1^-}),1-x(r^H_{11}+r^H_{1^- 1^-})) \, ,
\ee
where
\be
\begin{split}
O_L&=\frac{1}{\sqrt{2}} \begin{pmatrix} \sqrt{2} &\sqrt{2} x f_Q
r^H_{01} & -x f_Q\\ -2x f_Q r^H_{01} & 1+x(r^H_{1^- 1^-}-r^H_{11})
& -1+x(r^H_{1^- 1^-}-r^H_{11}) \\ 0 & 1-x(r^H_{1^- 1^-}-r^H_{11})
& 1+x(r^H_{1^- 1^-}-r^H_{11}) \end{pmatrix} \, , \\
O_R&=\frac{1}{\sqrt{2}} \begin{pmatrix}  \sqrt{2} &\sqrt{2} x f_d
r^H_{10} & x f_d\\ -2x f_d r^H_{10} & 1 & 1\\ 0 &1 &-1
\end{pmatrix} \, .
\end{split}
\ee
The interaction matrix of the quarks with the Higgs in the
interaction basis is
\be
\lambda=y^d \begin{pmatrix} f_Q f_d r^H_{00} & 0 & \sqrt{2} f_Q r^H_{01} \\
0 & 2 r^H_{1^- 1^-} & 0 \\ \sqrt{2} f_d r^H_{10} & 0 & 2 r^H_{11}
\end{pmatrix} \, ,
\ee
and in the quark-mass basis it is simply
$\lambda^{\mathrm{mass}}=O_L^{\dagger} \lambda O_R$.

The process that couples two opposite chirality zero-mode quarks
is carried out through the interaction of one quark with a heavy
mass eigenstate, a propagator that couples it to the opposite
chirality mass eigenstate and a coupling to the other light quark,
summing over the two heavy eigenstates. Specifically, the
effective coupling of the dipole amplitude is\footnote{~Here
$M_q^{\mathrm{mass}}$ should actually be divided by $M_{KK}$, to
avoid double-counting of the propagator with the calculation of
the previous appendix, and only consider the flavour structure
that the propagator introduces.}
\be
A \propto \lambda^{\mathrm{mass}}_{21}
\lambda^{\mathrm{mass}}_{12}/(M_q^{\mathrm{mass}})_{22}+
\lambda^{\mathrm{mass}}_{31}
\lambda^{\mathrm{mass}}_{13}/(M_q^{\mathrm{mass}})_{33} \, ,
\ee
which results into the overall coupling
\be \label{amp_flavor}
\frac{12v (y^d)^3 f_Q f_d r^H_{01} r^H_{10} r^H_{1^- 1^-}}{M_{KK}}\, .
\ee

\subsection{Coefficients of the Effective Operators}
\label{matching}

We are now ready to complete the calculation of the one-loop
dipole amplitudes and derive the coefficients of the
corresponding effective operators by a matching procedure.

In the case of the $s \rightarrow d\, g$ amplitude, the complete
result is obtained multiplying Eq.~\eqref{amp_no_flavor} and
Eq.~\eqref{amp_flavor}:
\be
A(s \rightarrow d\, g)=\frac{g_s t^a G^a_\mu}{4(4 \pi)^2 M_{KK}}
\overline{u}(p') \sigma^{\mu\nu}q_\nu (1\pm\gamma_5) u(p)
\frac{12v (y^d)^3 f_Q f_d r^H_{01} r^H_{10} r^H_{1^- 1^-}}{M_{KK}}
\, .
\ee
Inserting the appropriate projection operator, $\overline{u}(p')
\sigma^{\mu\nu}G_\mu q_\nu (1-\gamma_5) u(p)$ can be identified
with $\bar{s}_R \sigma^{\mu\nu} G_{\mu\nu} d_L$. Hence the
coefficient $C_G$ of the operator in Eq.~\eqref{chromo_operator}
is given by
\be \label{chromo_coefficient}
C_G =\frac{3}{16\pi^2 M_{KK}^2} f_{Q_1} (y^d)^3 f_{d_2} r^H_{01}
r^H_{10} r^H_{1^- 1^-} \, ,
\ee
and a similar expression is obtained for the opposite chirality
coefficient $C'_G$.

For $b \rightarrow s\gamma$, the entire calculation follows in the
same way. The coupling of a down-type KK quark to the photon adds
a factor of $1/3$. Moreover, there are two additional diagrams
with a charged Higgs and up-type quarks. We verified that our
calculation applies to one of these diagrams, with the photon
attached to the KK quark (with another factor of 2, for the charge
of an up-type quark relative to a down-type quark). The evaluation
of the diagram in which the photon is emitted by the charged Higgs
follows similarly. Here we simply use the result
of~\cite{Agashe:2008uz}, that the contribution of the latter is
$-1/6$ of the former. Note that this part contains an overlap
correction factor $r^H_{11}$ instead of $r^H_{1^- 1^-}$. Hence the
final result for the matching with the operator in
Eq.~\eqref{bsg_operator} is
\be \label{bsg_coefficient}
C'_7 =\frac{1}{4 \lambda_b M_{KK}^2} f_{Q_3} (y^u)^2 y^d f_{d_2}
r^H_{01} r^H_{10}(2r^H_{1^- 1^-} -\frac{1}{3} r^H_{11}) \,,
\ee
where in our actual numerical calculations we assume that $y^u$
and $y^d$ are of similar magnitude.

\end{document}